 \documentclass[aps,prl,twocolumn,showpacs,superscriptaddress]{revtex4}
\usepackage{graphicx}

\begin{document}

\title{ Spectral Correlation in Incommensurate Multi-Walled Carbon Nanotubes }

\author{K.-H. Ahn}
\affiliation{
School of Physics, Seoul National University, Seoul 151-747, Korea 
}
\author{Yong-Hyun Kim}
\affiliation{
Department of Physics, Korea Advanced Institute of Science and
Technology, Daejeon 305-701, Korea
}
\author{J. Wiersig}
\affiliation{
Max Planck Institute for Physics of Complex Systems, N\"othnitzer
Strasse 38, Dresden 01187, Germany
}
\author{K. J. Chang}
\affiliation{
Department of Physics, Korea Advanced Institute of Science and
Technology, Daejeon 305-701, Korea
}
\affiliation{
School of Physics, Korea Institute for Advanced Study,
Seoul 130-012, Korea
}

\date{\today}

\begin{abstract}
We investigate the energy spectra of clean incommensurate double-walled
carbon nanotubes, and find that the overall spectral properties are described
by the critical statistics similar to that known in the Anderson metal-insulator transition. 
In the energy spectra, there exist three different regimes characterized
by Wigner-Dyson, Poisson, and semi-Poisson distributions.
This feature implies that the electron transport in incommensurate
multi-walled nanotubes can be either diffusive, ballistic, or intermediate
between them,
depending on the position of the Fermi energy.
\end{abstract}

\maketitle

Carbon nanotubes have attracted great attention due to their remarkable
electrical and mechanical properties \cite{saito2}.
Many theoretical and experimental studies have demonstrated that single-walled
nanotubes (SWNTs) exhibit ballistic electron conduction \cite{tans1,bockrath},
and single-molecule devices utilizing semiconducting SWNTs have been
realized \cite{tans2}.
On the other hand, despite useful applications of multi-walled nanotubes
(MWNTs) \cite{tsukagoshi}, the transport properties of MWNTs are not well
understood and even controversial; conductance measurements using scanning
probe microscope showed ballistic behavior \cite{frank}, while
magnetoresistances measured for MWNTs on top of metallic gate indicated
diffusive conduction \cite{langer,bachtold}.

In general, MWNTs have very complex electronic structure, so that a direct
use of their transport properties is severely hindered.
If concentric carbon shells are especially incommensurate, the Bloch theorem
is no longer valid, and the Landauer formula for conductance is not applicable
because it is difficult to count the number of conducting channels \cite{datta}.
In fact, several experiments on MWNTs indicated that carbon shells have often
different periodicities \cite{iijima,ge}.
However, electronic structure calculations have been mostly focused for
commensurate multi-walled nanotubes so far.
Here, instead of directly calculating conductances, we investigate the
{\em spectral properties} of energy levels in incommensurate MWNTs.

The spectral analysis of energy levels has proven to be a useful tool to probe
the nature of eigenstates in disordered conductors \cite{montambaux}. 
In the diffusive metallic regime, the statistics of spectral fluctuations
is described by the Gaussian orthogonal ensemble (GOE)
of random-matrix theory \cite{mehta}.
In this regime, the distribution $P(s)$ of energy spacings between nearest
levels is well fitted by the Wigner-Dyson surmise 
$P_{\rm GOE}(s)=\frac{\pi}{2}s \exp(-\frac{\pi}{4}s^{2} )$,
where $s$ is in units of mean level spacing $\Delta$.
In the insulating regime, where the energy levels are uncorrelated, $P(s)$ is
given by the Poisson distribution, $P_P(s)=\exp(-s)$.
At the Anderson transition point, a critical statistics was found
\cite{shklovskii}. 
The short-range level correlation has been later described by semi-Poisson distribution
\cite{montambaux}, which has now been an important issue in quantum chaos \cite{hernandez,bogomolny,bogomolny2}.

In this Letter we make a spectral analysis of the energy levels of clean
incommensurate MWNTs, and find that the level statistics is similar to that
of the Anderson metal-insulator transition.
The energy spectra exhibit both the Wigner-Dyson and Poisson distributions
 depending on the
position of the Fermi level.
In addition, we demonstrate that incommensurate MWNTs have an intermediate
regime, which satisfies the semi-Poisson statistics.

\begin{figure}
\includegraphics[width=8.4cm]{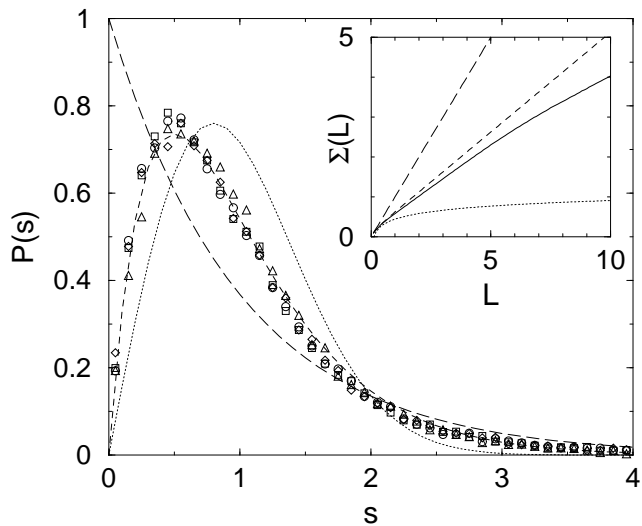}
\caption{The nearest energy spacing distribution $P(s)$  for
 energy levels between -7 and 7eV and different helicities 
(16,5)/(23,8) (triangle), (17,2)/(16,15) (circle), 
(16,5)/(17,15) (square), and (17,2)/(21,9) (diamond).
The semi-Poisson (dashed), Poisson
(long-dashed), and GOE (dotted) distributions were also plotted for comparison.
The inset shows the number variance $\Sigma(L)$ for (16,5)/(23,8) case.
The number of considered levels for (16,5)/(23,8) case is 20220. }
\label{fig:overall}
\end{figure}

To investigate the spectral correlation, we consider double-walled carbon
nanotubes (DWNTs), because the electrical conduction in MWNTs is believed
to be governed by the outermost shells \cite{bachtold,collins}.
We use a tight-binding model with one $\pi$-orbital per carbon atom,
which successfully describes the electronic structure of DWNTs
\cite{saito,roche}.
The tight-binding Hamiltonian is given by
\begin{eqnarray}
 H = \gamma_0  \sum_{i,j} c_{j}^{\dagger}c_{i}  - 
  W  \sum_{i^{\prime},j^{\prime}} \cos(\theta_{i^{\prime}j^{\prime}})
     e^{(a-d_{i^{\prime}j^{\prime}})/\delta}
     c_{j^{\prime}}^{\dagger}c_{i^{\prime}},
\label{hamiltonian}
\end{eqnarray}
where $\gamma_0$ (= -2.75 eV) is the hopping parameter between intra-layer
nearest neighbor sites, $i$ and $j$, and $W$ (= $\gamma_0$/8) is the strength
of inter-wall interactions between inter-layer sites, $i^{\prime}$ and
$j^{\prime}$, with the distance of $d_{i^{\prime}j^{\prime}}$ and the cut-off
for $d_{i^{\prime}j^{\prime}} > 3.9$ {\AA}.
Here $\theta_{ij}$ is the angle between two $\pi$ orbitals, $c_{i^{\prime}}$
is the annihilation operator of an electron on site $i^{\prime}$, 
$a$ (= 3.34 {\AA}) is the distance between two carbon walls, and
$\delta$ = 0.45 {\AA}.


To extract the fluctuations from the level sequence, it is customary to map
the real energy spectra $\{\epsilon_{i}\}$ onto the unfolded spectra
$\{E_{i}\}$ through $E_{i}=\bar{N}(\epsilon_{i})$, where $N(\epsilon_{i})$
is the number of levels up to $\epsilon_{i}$ and the overline denotes its
broadened value \cite{haake}.
Here we use the Gaussian broadening scheme for averaging the energies
\cite{GMRR}. 
After unfolding, we obtain the distribution $P(s)$ as a function of
nearest energy spacing, $s_{i}=E_{i}-E_{i-1}$ (the mean level spacing of
the unfolded spectra equals 1), and the number variance
$\Sigma(L)=\big< (n(L,E)-L )^{2} \big>$, which represents the variance
of the number of levels in the interval $(E-L/2,E+L/2)$, i.e.,
$n(L,E) = N(E+L/2)-N(E-L/2)$.
The bracket $\big<\cdots\big>$ denotes the average around the energy $E$
on an energy window much larger than the mean level spacing but much smaller
than $E$.

The two shells of DWNTs are incommensurate if the ratio
of the unit cell lengths along the tube axis is irrational.
We test many incommensurate DWNTs, where both the shells
are metallic or semiconducting, a semiconducting shell is inside a metallic
tube, and vice versa.
The diameters of the inner and outer tubes are 1.5 and 2.2 nm, respectively,
and the nanotube length is set to be about 53 nm.
We find the spectral properties are  similar for all the incommensurate
DWNTs with different helicities.  (See the $P(s)$ in Figure \ref{fig:overall}).
Since the spectral properties are similar for all the tubes considered here,
from here on, we only present the spectral properties of the (16,5)/(23,8) DWNT
where the semiconducting (16,5) single-walled tube is aligned inside
the metallic (23,8) tube.
In the inset of Figure \ref{fig:overall}, $\Sigma(L)$ is also shown
for the (16,5)/(23,8) nanotube. 
We find that $P(s)$ and $\Sigma(L)$ cannot be described by the Poisson
distribution or GOE, but, they are well described by the semi-Poisson (SP)
distribution \cite{hernandez,bogomolny,bogomolny2}; 
\begin{eqnarray}
P_{SP}(s)&=&4s\exp(-2s) \\
\Sigma_{SP}(L)&=&\frac{L}{2}+\frac{1}{8}(1-e^{-4L}).
\end{eqnarray}
The SP distribution is defined by removing every other level from an ordered
Poisson sequence and turned out to be a reference point for the critical statistics
of several disordered systems like the Anderson model at mobility edge
\cite{shklovskii,braun}, and also of other systems without disorder such as
pseudo-integrable quantum billiards \cite{bogomolny,bogomolny2,wiersig}.

While $P(s)$ carries information on short-range correlation in the energy
spectra, $\Sigma(L)$ contains rather long-range correlation.
Usually $\Sigma(L)$ indicates a deviation from the universal value for large
$L$, thus, the particle dynamics becomes nonuniversal at short time scale. 
For large $L$, the spectral correlation is not universally semi-Poisson
but depends on the helicity of nanotubes. 
The deviation from the SP distribution gives useful information;
the linear behavior for large $L$, i.e., $\Sigma(L)/L \rightarrow \chi$,
represents the level compressibility, and for disordered metals
$\chi$ was shown to be \cite{chalker1}
\begin{eqnarray}
\chi= \frac{1}{2}\big(1- \frac{D_{2}}{d}\big),
\label{chi}
\end{eqnarray}
where $D_{2}$ is the multifractal exponent of the inverse participation
ratio and $d$ is the spatial dimension.
Here $D_{2}/d$ is related to the 'probability of return' \cite{chalker2},
\begin{eqnarray}
p(t) \propto t^{-D_{2}/d}.
\label{preturn}
\end{eqnarray}
Our incommensurate DWNT behave as a two-dimensional systems ($d=2$)
with off-diagonal disorder, which has a mixed boundary between periodic
and hard-wall conditions. 
From the inset in Fig.~\ref{fig:overall}, we estimate $D_{2}/d$ to be
about 0.32 for the DWNT considered here.
While $p(t)\propto t^{-d/2}$ in normal diffusion,
our system exhibits anomalous diffusion, which was also found by previous 
wavepacket spreading calculations \cite{roche}.

The SP distribution in Fig.~\ref{fig:overall} is mostly contributed
from the region of large $|E|$, where the density of states (DOS) is high,
while the energy statistics is usually not independent of the energy regime.
Since real electron conductions occur near the Fermi level, it is 
instructive to examine the energy-level statistics on various energy
windows.
Although we discuss the level statistics when the Fermi level increases,
we find similar statistics for the downward shift of the Fermi level,
which usually occurs in hole-doped tubes.

\begin{figure}
\includegraphics[width=8.5cm]{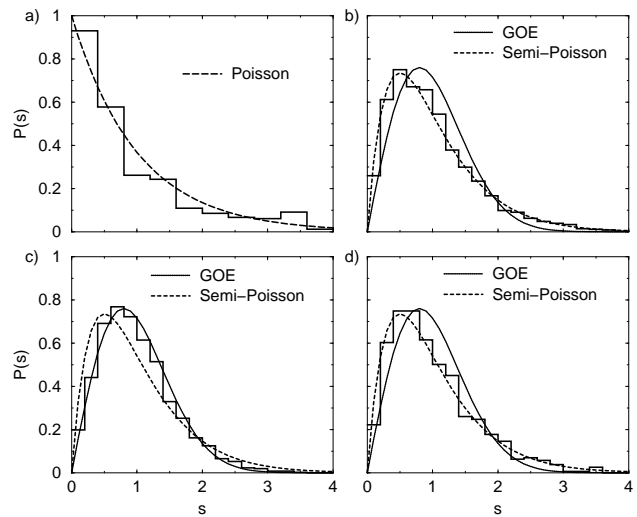}
\caption{ The nearest energy spacing distribution $P(s)$ on the energy windows
of (a) (0.08,1.3eV) with 511 levels, (b) (1.0,2.5eV) with 1872 levels,
(c) (2.5,3.5eV) with 2696 levels, and (d) (3.5,4.0eV) with 888 levels. }
\label{fig:windows}
\end{figure}

In the level statistics, we remove the states lying between 0 and 0.08 eV,
which are associated with the localized states near the sample boundary
due to the finite size of the system.  
In Fig. \ref{fig:windows}(a), $P(s)$ is drawn for low energy states
from 0.08 to 1.3 eV, and exhibits the Poisson distribution, indicating 
that the energy levels are uncorrelated.
Single-wall nanotubes, which are constituents of the DWNT, 
are approximately described in terms of $P_{P}(s)$ and $\delta(s)$.
In this case, the delta-function peak results from the doubly degenerate
states due to inversion symmetry.
If inter-wall interactions are absent in the DWNT, 
a superposition of two independent spectra of the nanotube constituents
is also described by the Poisson and delta functions. 

When inter-wall interactions break the degeneracy, smearing out the
delta-function peak, $P(s)$ is still Poisson-like, as shown in
Fig. 2(a), implying that the energy mixing between the two carbon shells
is insignificant for low energies.
In the low energy regime, it is useful to use the Landauer formula,
\begin{eqnarray}
G=\frac{2e^{2}}{h}\sum_{i}T_{i}(E_{F}),
\end{eqnarray}
where $T_{i}(E_{F})$ is the transmission probability of the channel
$i$ obtained from  the band structure calculations at the Fermi energy $E_{F}$ \cite{datta}.
Our results infer that quantized conductances observed in MWNTs \cite{frank}
may be due to the fact that the Fermi level is close to the charge
neutrality point, $E_{F} \approx 0$.
However, it is usually very difficult to find experimentally the location
of the Fermi level.

As the Fermi level is shifted further, since 
the mean level spacing becomes smaller and the inter-wall interaction matrix elements become more significant,
the inter-wall interactions induce
a more significant mixing of the energy levels between the carbon shells.
For higher energy states, we find that $P(s)$ follows the SP distribution,
as shown in Figs. \ref{fig:windows}(b) and (d).
The quantum conductance in this regime still remains a challenging
problem. 
In particular, it is a nontrivial but important task to find the tube length dependence of
quantum conductance in this regime.
Recent calculations \cite{yoon} showed negligible inter-wall tunneling currents
for the tube length of 1 $\mu$m, but it is questionable whether one can use the Bardeen's formalism for DWNTs where 
the coherent back-tunneling between carbon shells is non-negligible. 
According to the weak localization formula, 
$\delta \sigma \propto -\int dt p(t)$, with Eq. (\ref{preturn}), conductances may depend on the
tube length. 
However, the weak localization formula considers only the interference of the pair of time-reversal paths, which is not
guaranteed in the critical regime.

In Fig. \ref{fig:windows}(c), one can see that the critical statistics evolves
to the GOE as going to the higher energy region, where $P(s)$ is quite close to
the Wigner-Dyson surmise. 
We also find the GOE-like statistics for both $P(s)$ and $\Sigma(L)$ in all
the incommensurate DWNTs tested here.
The appearance of the GOE-like behavior indicates that all the symmetries
except for time-reversal symmetry are effectively broken,
 and the corresponding electron motion is ergodic.
 In this regime, the conductivity of MWNTs
may be described by the weak localization theory.
One should also note that the weak localization theory is applicable
for energies higher than those in the first critical regime, because the sequence of
statistics with increasing energy is
 $\text{Poisson} \rightarrow \text{SP} \rightarrow \text{GOE} \rightarrow \text{SP}$ .

In MWNTs on top of metallic gates, the details of contacts, local excessive
charges, and depletion of charge carriers may change the position of the
Fermi level \cite{xue}. 
Based on our results, we guess that in conductance measurements
\cite{tans1,tans2,bachtold}, the Fermi level is shifted to the GOE regime
or to the critical regime (close to the GOE) where conductivity is properly described by the weak
localization formula.

\begin{figure}
\includegraphics[width=8.4cm]{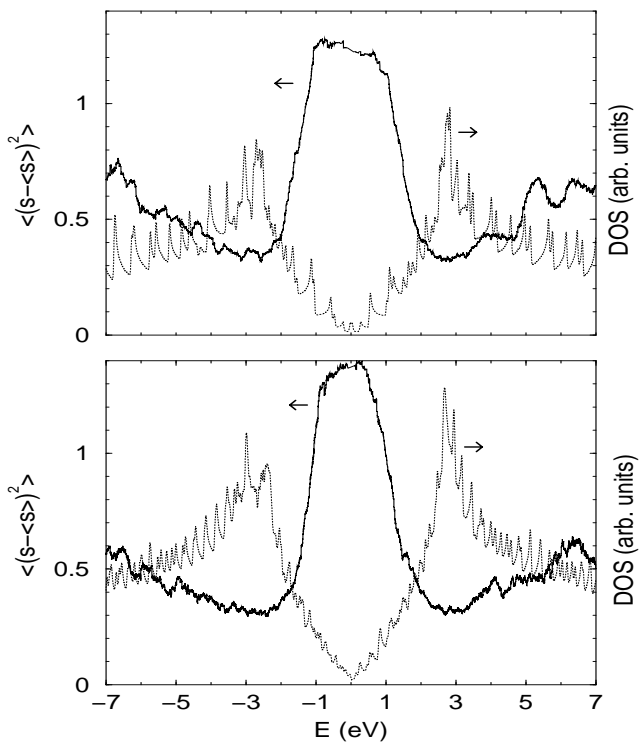}
\caption{The variance of the level spacing (solid) in Eq.~(\ref{eq:vars})
and the density of states (dotted) in arbitrary units for the (16,5)/(23,8)
(upper panel) and (43,8)/(51,9) (lower panel) double-walled tubes. }
\label{fig:vars}
\end{figure}

The calculated variances over every 1000 consecutive levels are compared
with the densities of states for two different DWNTs in Fig. \ref{fig:vars}.
The size of fluctuations of the level spacing is dictated by the spectral
statistics; 
\begin{eqnarray}
\label{eq:vars} && \big< (s- \big< s \big>)^{2} \big>
 = \int_{0}^{\infty} ds (s-1)^{2}P(s) 
\\ \nonumber &=& 1, ~ \frac{1}{2},~ \frac{4}{\pi}-1
~~({\rm Poisson,~~SP,~~GOE}). 
\end{eqnarray}
It is expected that the effect of inter-wall interactions on the level
spacing is most significant when the mean level spacing $\Delta$ is
minimum near $E=\pm \gamma_{0}=\pm 2.75eV$, where the maximum DOS occurs
on a graphene sheet.
One can see
 $\big< (s- \big< s \big>)^{2} \big>=\frac{4}{\pi}-1 \approx 0.27 $ near 
$E=\pm \gamma_{0}=\pm 2.75eV$, which is a signature of GOE distribution.
The variances larger than 1 near $E =0$ are due to the fact that
$P(s)$ is not completely Poisson-like because the degenerate levels of
each nanotube layer are not completely lifted up.

We find that a transition from the Poisson-like to SP-like statistics
occurs at lower energies as the diameter of nanotubes increases. 
One of the reasons for this trend seems that the DOS is enhanced
for nanotubes with larger diameters.
The transition energy is close to 1 eV in the (43,8)/(51,9) DWNT with
the outer and inner diameters of 4.4 and 3.7 nm, respectively,
while it is higher than 1 eV for the (16,5)/(23,8) tube with the diameters of
2.2 and 1.5 nm, as shown in Fig. \ref{fig:vars}.
Since experimentally measured diameters of MWNTs are often in the range of
10 nm, we expect that the crossover of spectral statistics occurs
at energies lower than 1 eV.

Very recently, {\it Kociak} and his co-workers\cite{kociak} 
measured the conductance of a DWNT whose two tubes have a gap and showed 
linear conductance near the Fermi level.
The linear conductance or a finite DOS near the Fermi level indicates
that the Fermi level indeed can be shifted far from the charge neutral point. 

Finally, we point out that the spectral statistics might be probed
through conductance measurements in the Coulomb blockade regime\cite{buitelaar,woodside}.
The temperature should be low enough to ensure that
$k_{B}T < U, \Delta $, where $U$ is the charging energy of nanotube dot.
In the resonant tunneling regime, where electron tunnels through one
quantum level, if electron-electron interactions are not too strong
\cite{ahn}, the conductance peak spacing with varying the gate voltage
provides information on the single-particle energy spacing.

In conclusion, we have shown that the spectral statistics of incommensurate
double-walled nanotubes follows the Poisson, GOE, or SP distribution,
depending on the energy window, while the overall states are well described
by the SP distribution.
This results indicate that the nature of electron transport in multi-walled
nanotubes can be either ballistic, diffusive, or critical, depending on
the position of the Fermi level.
It is questioned whether the usual weak localization correction is relevant to
existing experiments. 
The Coulomb blockade oscillation in nanotube dots is suggested to investigate 
the spectral statistics in this work.

\begin{acknowledgments}
The authors acknowledge J. Ihm, J. Yu, B. L. Altshuler, G. Montambaux, 
K. Richter, S. Evangelou, H.-S. Sim, G. Cuniberti, and H.-W. Lee 
for useful discussions.
This work was supported by QSRC and Brain Korea 21 project.
\end{acknowledgments}

\newpage

\end{document}